\documentclass[lettersize,journal]{IEEEtran}
\usepackage{amsmath,amssymb,amsfonts}
\usepackage{algorithmic}
\usepackage{algorithm}
\usepackage{array}
\usepackage[caption=false,font=normalsize,labelfont=sf,textfont=sf]{subfig}
\usepackage{textcomp}
\usepackage{stfloats}
\usepackage{url}
\usepackage{verbatim}
\usepackage{graphicx}
\usepackage{cite}
\usepackage{braket}

\begin{document}

\title{Charge Parity Rates in Transmon Qubits with Different Shunting Capacitors}

\author{Yi-Hsiang Huang, Haozhi Wang, Yizhou Huang, Sylvie McKnight-Milles, Zachary Steffen, and B. S. Palmer
\thanks{The authors are with the Laboratory for Physical Sciences and the University of Maryland, College Park, MD 20740 USA}}


\maketitle

\begin{abstract}
The presence of non-equilibrium quasiparticles in superconducting resonators and qubits operating at millikelvin temperature has been known for decades. One metric for the number of quasiparticles affecting qubits is the rate of single-electron change in charge on the qubit island ({\it i.e.} the charge parity rate). Here, we have utilized a Ramsey-like pulse sequence to monitor changes in the parity states of five transmon qubits. The five qubits have shunting capacitors with two different geometries and fabricated from both Al and Ta. The charge parity rate differs by a factor of two for the two transmon designs studied here but does not depend on the material of the shunting capacitor. The underlying mechanism of the source of parity switching is further investigated in one of the qubit devices by increasing the quasiparticle trapping rate using induced vortices in the electrodes of the device. The charge parity rate exhibited a weak dependence on the quasiparticle trapping rate, indicating that the main source of charge parity events is from the production of quasiparticles across the Josephson junction. To estimate this source of quasiparticle production, we simulate and estimate pair-breaking photon absorption rates for our two qubit geometries and find a similar factor of two in the absorption rate for a background blackbody  radiation temperature of $T^*\sim$ 350 mK. 
\end{abstract}


\section{Introduction}
\IEEEPARstart{A}{t} temperatures below 50 mK the expected thermal density of quasiparticle (QP) excitations for aluminum should be less than $n_{qp} < 1\ \textrm{cm}^{-3}$. Instead, superconducting qubit experiments have found densities on the order of  $n_{qp} \sim (10^{-2}$ to $ 1)\  \mu m^{-3}$ \cite{2013Riste,2014Wang,2018Serniak,2020Vepsalainen,2022Diamond,2022Pan,2024Connolly,2024Harrington,2024Liu}. Understanding the source and dynamics of these quasiparticles is crucial to evaluate and improve the performance of the devices. For superconducting transmon qubits, a quasiparticle tunneling across the Josephson junction can change the charge parity by a single electron, induce energy relaxation, and be a source of dephasing. Infrared photons \cite{2011Barends,2011Córcoles}, cosmic rays \cite{2021Wilen,2021Cardani,2024Harrington}, and radioactive isotopes \cite{2020Vepsalainen,2021Cardani,2023Cardani,2024Loer} have been found or suggested to be possible sources of non-equilibrium  quasiparticles. 

One quantifiable metric for the amount of quasiparticles in a qubit system is measurements in real time of changes in charge by a single electron or the charge parity rate ($\Gamma_{e\leftrightarrow o}$) \cite{2013Riste,2014Catelani,2018Serniak,2019Serniak,2021Glazman,2022Gordon,2022Diamond,2022Iaia,2022Kurter,2022Pan,2024Connolly,2024Rafsanjani}. In these measurements, a transmon qubit is fabricated to have a resolvable amount of charge dispersion by making the ratio of the Josephson energy to single-electron charging energy $E_{J}/E_{c}\sim 20-30$. By monitoring the qubit frequency with a fast sampling rate, typically using a modified Ramsey coherence scheme, changes in the charge parity can be discerned.  Recently, a new mechanism that directly links a QP generation event to the change in charge parity was proposed by Houzet {\it et al.}: a photon with energy greater than twice the superconducting gap (2$\Delta$) is absorbed at the qubit junction resulting in a QP on both sides of the junction \cite{2019Houzet,2022Diamond,2024Connolly}. Furthermore, Liu {\it et al.} suggested that the absorption of pair-breaking photons by the Josephson junction is facilitated by unwanted antenna modes parasitic to the pads forming the shunting capacitor in a superconducting qubit \cite{2021Rafferty,2022Pan,2024Liu}.
\par
 In this work, we use a Ramsey-like pulse sequence to map the charge parity state of five different transmon qubits. The qubits measured had two different designs:  an ``x-mon'' geometry galvanically connected to the large ground plane of the device and a ``two-pads'' geometry galvanically isolated from the ground plane. We also measured devices with shunting capacitors made from aluminum and tantalum. For the shunting capacitors made from aluminum,  the superconducting gap ($\Delta_{Al,\dashv \, \vdash}$) is nominally smaller than the gap of the two electrodes ($\Delta_{Al,-\mkern-8mu\times\mkern-8mu-}$) that form the Al/AlOx/Al junction. Meanwhile for the Ta shunting capacitor,  $\Delta_{Ta,\dashv \, \vdash}$ is larger than $\Delta_{Al,-\mkern-8mu\times\mkern-8mu-}$ of the two Al electrodes. 

 In order to further investigate the origin of the charge parity switching, we intentionally introduce magnetic vortices in one of the Al transmons to change the trapping rate of  quasiparticles \cite{2014Nsanzineza,2014Wang} and measure the charge parity rates. From these results we conclude that the dominant source of charge parity events is from the creation of QPs at the junction. Finally, we perform simulations and model the charge parity rate of the two geometries of transmons measured here to extract an efficiency of quasiparticle generation and an effective  temperature to explain the observed parity rates.

\section{Experimental Methods}
\begin{figure}
    \centering
    \includegraphics[width=\linewidth]{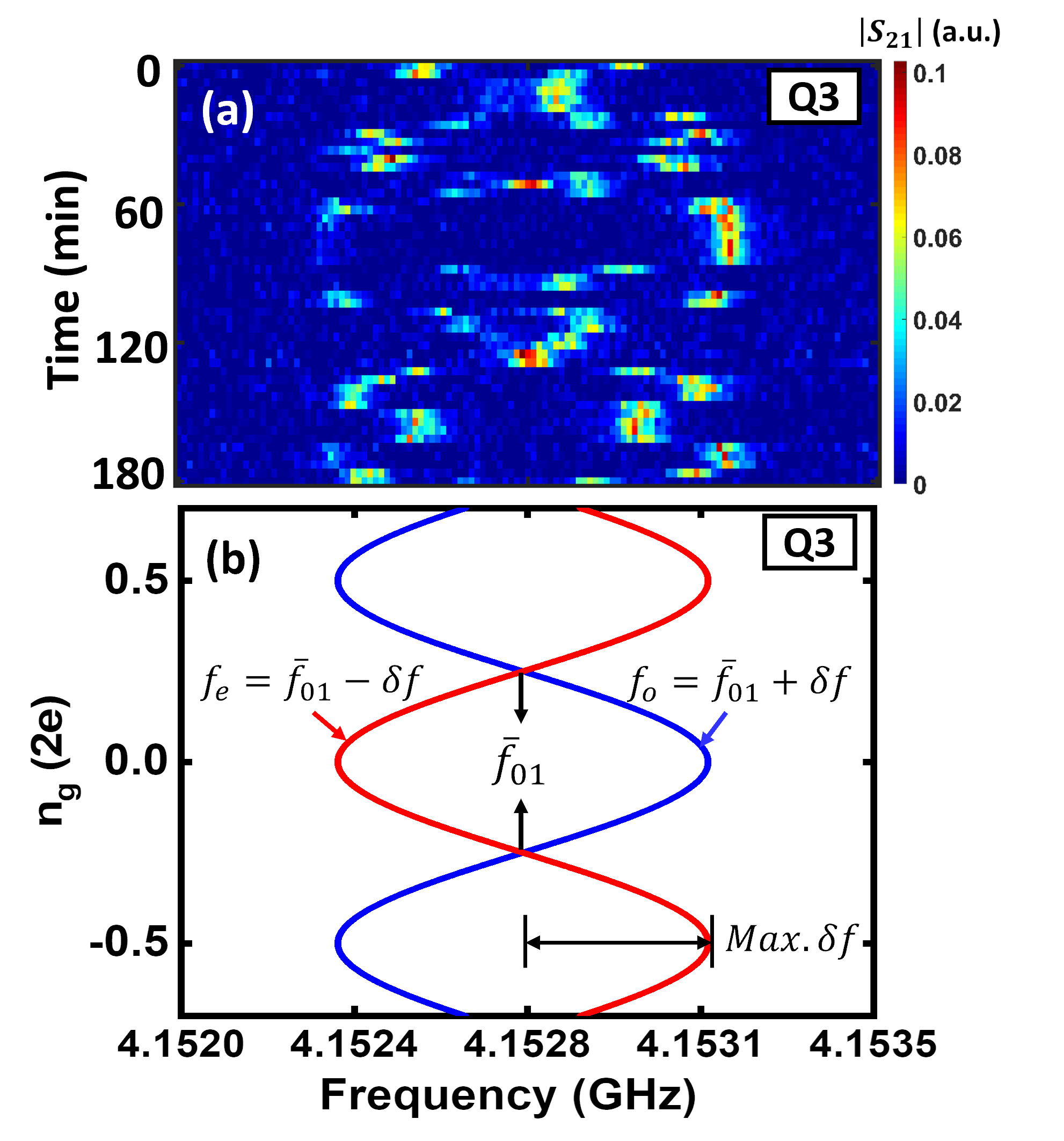}
    \caption{(a) Spectroscopic measurements of the first excited state transition $f_{01}$ of a charge sensitive transmon (Q3) over a three hour duration. (b) The predicted qubit ground to first excited state transition frequency versus offset gate charge ($n_{g}$). Away from $n_g = \pm 0.25$, two $f_{01}$'s centered around $\Bar{f}_{01} = 4.1528$ GHz are measured in (a),  corresponding to the charge parity even (red) and odd (blue) state on the superconducting island of the qubit or vice versa. The $n_g$-dependent $\delta f$ drifts over a 3-hour measurement period as a result of uncontrollable changes in $n_g$.}
    \label{fig:1}
\end{figure}

Charge parity rates were measured nominally at $T=20$ mK in five offset-charge-sensitive transmon qubits with $18 \leq E_J/E_c \leq 32$ and with transition frequencies 3.41 GHz $\leq f_{01} \leq$ 4.5 GHz (Table \ref{tab:1}). Each transmon was coupled to its own readout resonator which had resonant frequencies from 6.66 GHz $\leq f^{bare}_{r} \leq$ 7.29 GHz, enabling high-fidelity dispersive readout. Transmons with these parameters exhibit a measurable amount of charge dispersion, such that the qubit ground to first excited state transition frequency drifts over many hours due to uncontrollable offset charge drift (see Fig. \ref{fig:1}a). Nominally, away from a reduced offset gate charge $n_g$ (defined with respect to Cooper-pairs of 2{\it e}) of $\pm 0.25$, two $f_{01}$'s centered around $\Bar{f}_{01}$, $f_{01} = \Bar{f}_{01} \pm \delta f$, are observed in spectroscopic measurements. From extraction of the qubit parameters for each device, $2\delta f$ corresponds to changes in the charge of the island of the transmon by one electron therefore changing the parity of the device from odd to even or vice versa. We denote frequencies associated with the two charge states as $f_o$ and  $f_e$ (Fig. \ref{fig:1}b). The questions being addressed here is what is the noise mechanism for the source of these changes in the parity states and  what is the rate of $1e$ charge parity changes? 

\begin{figure*}[h!]
    \centering
    \includegraphics[width=\linewidth]{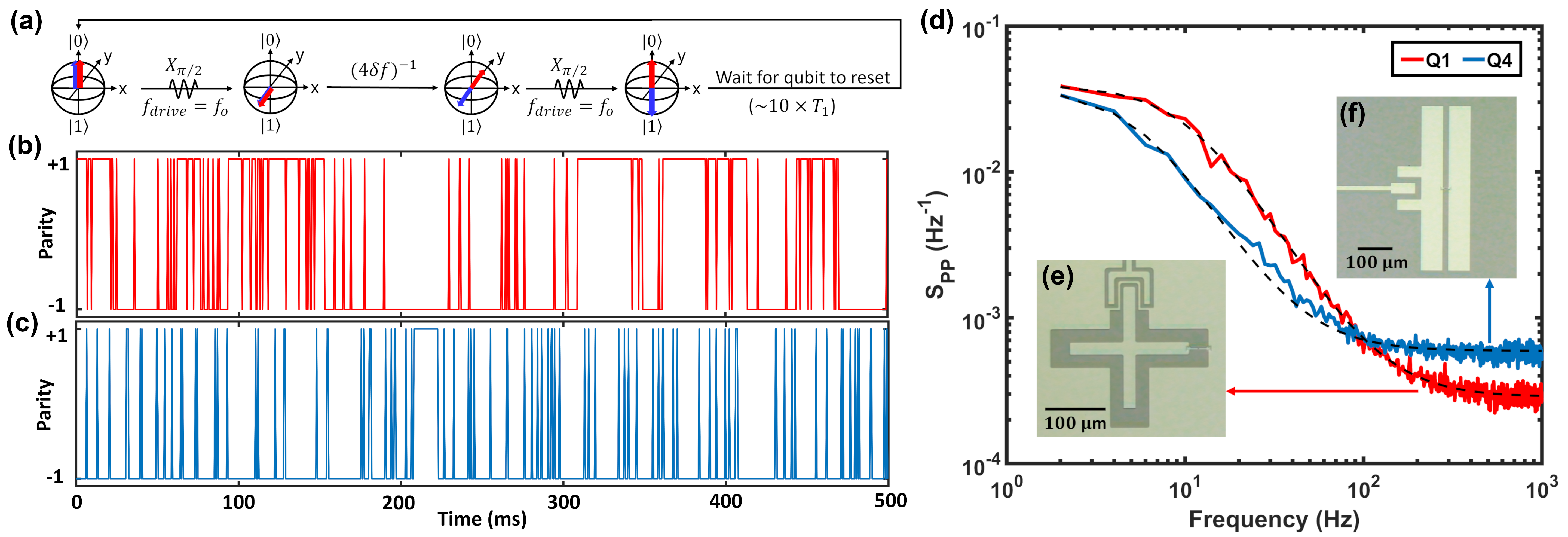}
    \caption{Probing the charge parity state of a qubit in real time. (a) A Ramsey-like pulse sequence is used to map out the charge parity state. A waiting time of $10 \times T_1$ at the end of each cycle allows the qubit to relax to $\ket{0}$ before the start of next cycle. Sample parity vs. time traces measured over a total interval of 500 ms for (b) x-mon style Q1 and (c) two-pads style Q4. (d) The power spectral density ($S_{PP}$) of the charge parity states. Optical images showing the shunting capacitor designs for (e) x-mon and (f) galvanically isolated two-pads.}
    \label{fig:2}
\end{figure*}

To study the source of charge parity rates, the five qubits measured had different geometries for their shunting capacitor and were fabricated from different superconducting gap material.
Table \ref{tab:1} lists the qubits measured in this experiment and their parameters. Q1-Q3 had an ``x-mon'' geometry, where one side of the shunting capacitor is galvanically connected to the ground plane of the device (Fig. \ref{fig:2}e). Q4 and Q5 were galvanically isolated with a ``two-pads'' geometry (Fig. \ref{fig:2}f). Q1-Q4 had shunting capacitors made of 200 nm thick aluminum with a $T_c= 1.1$ K corresponding to a zero temperature gap of $\Delta_{Al,\dashv \, \vdash} = 170 \, \mu$eV, while Q5's shunting capacitor was fabricated from 200 nm thick  tantalum with a $T_c= 4.3$ K (see Device Fabrication for details). Based on previous measurements, we nominally expect the Al junction electrodes, with thicknesses ranging from 30 $\sim$ 160 nm (see Table~\ref{tab:1}), to have a superconducting gap as large as $\Delta_{Al,-\mkern-8mu\times\mkern-8mu-} \simeq 200\, \mu$eV. From differences in the gaps between the shunting capacitors and junction electrodes, we  expect any quasiparticles generated in the Al based transmons to accumulate in the shunting capacitor \cite{2022Pan}. On the other hand, the Ta material should be less susceptible to pair breaking from blackbody radiation but any quasiparticles generated in the device would be expected to accumulate in the Al electrodes of the Josephson junction. Furthermore, for the Al junction electrodes, a larger difference  in $\Delta_{Al,-\mkern-8mu\times\mkern-8mu-}$ is created when the difference in the thicknesses of the two junction electrodes is larger, resulting in an expected smaller $\Gamma_{e\leftrightarrow o}$ when QPs are already present in the electrodes \cite{2022Marchegiani,2023Steffen,2024Krause}.

\begin{table*}[b]
    \caption{Summary of important qubit parameters and measured parity rate $\Gamma_{e \leftrightarrow o}$.}
    \label{tab:1}
    \centering
    \begin{tabular}{|c|c|c|c|c|c|}
        \hline
        Device No. & Q1 & Q2 & Q3 & Q4 & Q5\\\hline
        Materials: Shunting Capacitor / Substrate  & \multicolumn{4}{c|} {Aluminum / Silicon}  & Tantalum / Sapphire \\\hline
        Junction Fabrication Style \& Two Thicknesses (nm) & \multicolumn{4}{c|}{Dolan 30 \slash 160}  & Manhattan 35 \slash 70 \\\hline
        Geometry & x-mon & x-mon & x-mon & two-pads & two-pads\\\hline
        $E_J/E_c$ & 30 & 32 & 28 & 31 & 18\\\hline
        Average $\Bar{f}_{01}$ (GHz) & 4.38 & 4.50 & 4.15 & 4.32 & 3.41\\\hline
        Maximum $\delta f$ (MHz) & 0.24 & 0.17 & 0.40 & 0.22 & 4.30\\\hline
        $T_1$ ($\mu$s) & 51 & 57 & 94 & 69 & 102\\\hline
        $\Gamma_{e \leftrightarrow o}$ (s$^{-1})$ & 33.3 & 37.0 & 33.3 & 17.5 & 17.2\\\hline
    \end{tabular}
\end{table*}

Prior to measuring the charge parity rates, a Ramsey measurement at different delay times between the two $\pi/2$ qubit pulses was conducted providing information of $\delta f$, a quantity that  slowly changes due to drift in the offset gate charge. After determining $\delta f$, a $\pi/2$ pulse at $f_{drive}=f_o$ was used to excite the qubit from the ground state $\ket{0}$ to a superposition state $(\ket{0}+\ket{1})/\sqrt{2}$. Depending on whether the island of the qubit is in the odd or even charge state and by waiting a time interval $(4 \delta f)^{-1}$, the state of the qubit accrues a phase difference of $\pi$ due to differences in the precession frequency. A final $\pi/2$ pulse was then applied at $f_{drive}=f_o$  driving the qubit to the $\ket{0}$ state, when the qubit island has the even parity, or $\ket{1}$ state, when the qubit island has the odd parity. A single-shot measurement is then performed with a characterized fidelity  between $\mathcal{F}=$ 0.85 to 0.95 for the different qubit-resonator devices, enabling the discernment of the $\ket{0}$ from the $\ket{1}$ state. We then wait a duration of $10 \times T_{1}$ to allow the qubit to relax to the ground state before the measurement is repeated again (see Fig.~\ref{fig:2}(a) for the charge parity sequence). Using this technique, the charge parity state was sampled at a time interval of 1 ms, a time rate much faster than spectroscopic measurements, and over a time duration of 0.5 sec (see Fig.~\ref{fig:2} (b) and (c) for sample traces). 

\section{Results and Discussion}
From the acquired charge parity time traces, the power spectral density of charge parity states $S_{PP}$ was calculated. Since the raw time trace data is binned into two states, a telegraph noise model is used to extract the charge parity rate $\Gamma_{e \leftrightarrow o}$ \cite{2013Riste,2018Serniak,2022Pan}:
\begin{equation}
S_{PP}=\frac{A\Gamma_{e \leftrightarrow o}}{(\Gamma_{e \leftrightarrow o})^2+(\pi f)^2}+C.
\end{equation}
Here a constant $C$ is included to account for the sampling noise due to infidelity of the measurement. Fig. \ref{fig:2}d shows the corresponding $S_{PP}$ for the time trace data in Fig.~\ref{fig:2}(b) and (c).

Table~\ref{tab:1} lists the extracted $\Gamma_{e \leftrightarrow o}$ for all of the measured devices. First, we note that the measured values of $\Gamma_{e \leftrightarrow o}$ are in line with the previously reported $\Gamma_{e \leftrightarrow o} \approx (10^{-1} \sim 10^{5}) \, s^{-1}$ \cite{2013Riste,2018Serniak,2019Serniak,2022Gordon,2022Diamond,2022Kurter,2022Iaia,2022Pan,2024Connolly} where the lower value in the rates was achieved with IR filters in the microwave lines, which were not used in this experiment. Second, we note that the measured $\Gamma_{e \leftrightarrow o}$  for the x-mon devices are approximately twice as large than the two-pads geometry. Third, $\Gamma_{e \leftrightarrow o}$ for Q4, an Al based device, and Q5, a Ta based device, were nearly the same. These observations suggested to us that the observed parity events are formed from pair breaking events across the junction\cite{2019Houzet}.

To further test this conjecture, vortices were intentionally introduced  in the Q3 device and $\Gamma_{e \leftrightarrow o}$ was remeasured. The idea for this experiment is that if quasiparticles are formed in the electrodes then the presence of vortices  would have the tendency to trap  quasiparticles, decrease the background quasiparticle density, prevent them from tunneling across the junction, and therefore decrease $\Gamma_{e \leftrightarrow o}$. Vortices were introduced in the device by cooling the device through its $T_c$ in a magnetic field perpendicular to the film of the device (see Measurement Setup for more details).

\begin{figure}
    \centering
    \includegraphics[width=\linewidth]{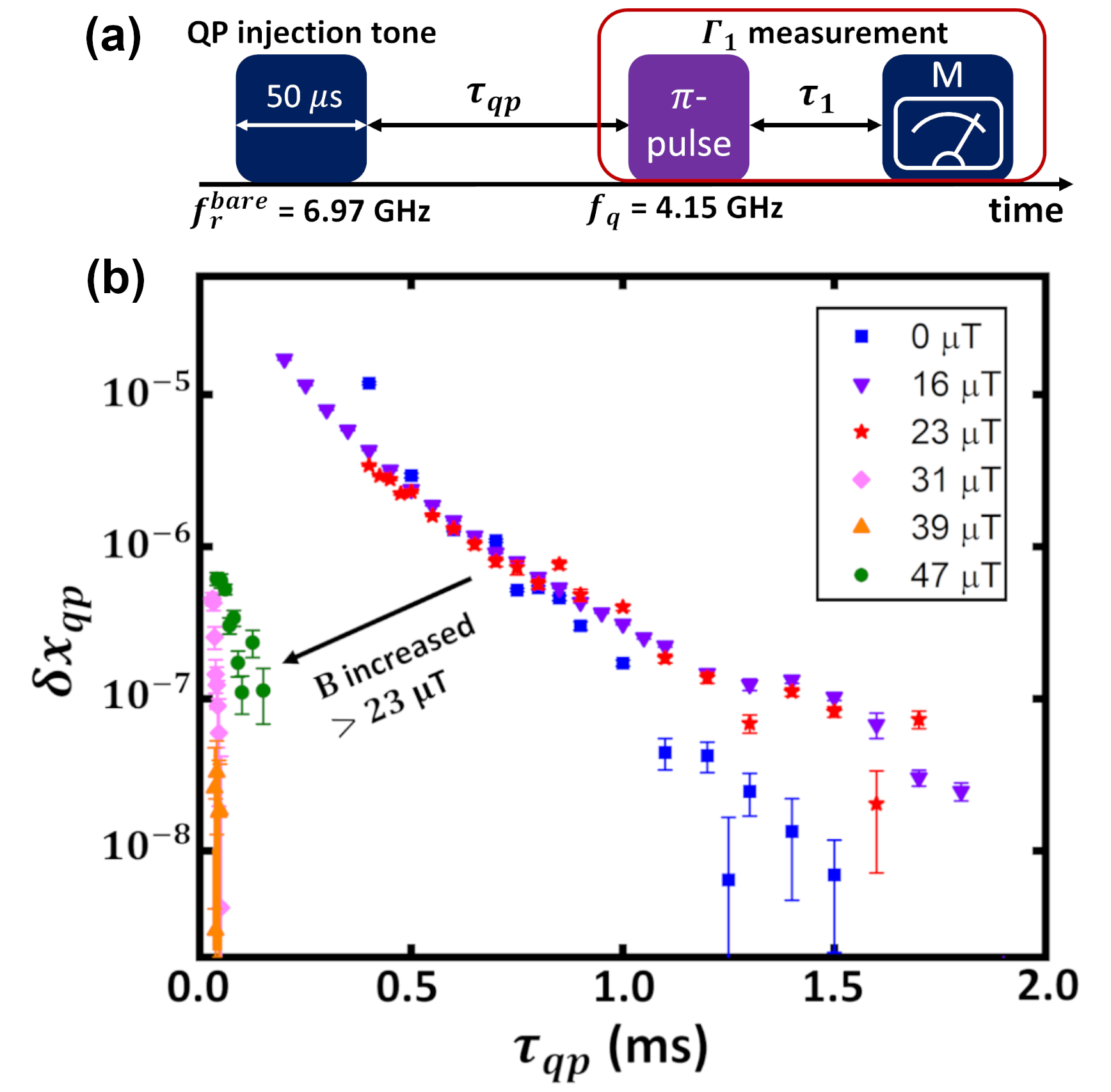}
    \caption{Measurement of QP injection and recovery under various applied magnetic field $\bf B$ on device Q3. (a) Pulse sequence used to study excess QP dynamics. First, excess QPs are intentionally generated at the junction by sending a microwave tone at $f_r^{bare}$. After a delay time $\tau_{qp}$, the additional decay rate $\Delta \Gamma_1$ is measured and related to $\delta x_{qp}$. (b) Semi-log plot of $\delta x_{qp}$ as a function of $\tau_{qp}$ under various $\bf B$.}
    \label{fig:3}
\end{figure}

The effect of the magnetic field on the quasiparticle trapping rate was first examined by deliberately injecting excess QPs and measuring the decay of these QPs due to recombination and trapping \cite{2014Nsanzineza,2014Wang}. Fig. \ref{fig:3}a shows the pulse sequence used to study excess QP dynamics. First, a microwave tone is sent at the bare resonator frequency $f_r^{bare}$ = 6.97 GHz with enough amplitude to both excite the resonator and induce an ac voltage $V > 2 \Delta_{Al,-\mkern-8mu\times\mkern-8mu-} \, / \, e$ across the junction of the qubit. This large  voltage across the junction results in quasiparticle-quasiparticle tunneling and therefore excess quasiparticles in the two electrodes\cite{2014Wang,2017Patel,2022Iaia,2023Bargerbos}. After a quasiparticle recovery delay time $\tau_{qp}$, the qubit's energy decay rate $\Gamma_1$ is measured  by sending a $\pi$ pulse to the qubit and introducing a secondary varying delay time $\tau_1$ before the qubit state is  measured. For short $\tau_{qp}$, $\Gamma_1$ is large, due to a large amount of excess quasiparticles present in the system. Using a symmetric gap model, the excess normalized QP density $\delta x_{qp}$ is related to the  excess qubit decay rate\cite{2009Martinis,2014Wang,2019Houzet,2020Vepsalainen}
\begin{equation}
\delta x_{qp} = \sqrt{\frac{\pi \hbar}{4f_q \Delta_{Al,-\mkern-8mu\times\mkern-8mu-}}} \Delta \Gamma_1.
\end{equation}

Fig. \ref{fig:3}b shows the extracted $\delta x_{qp}$ versus the separation $\tau_{qp}$ at different applied magnetic fields ($\bf B$). Above ${\bf B}$ = 23 $\mu$T, the excess quasiparticle recovery time is observed  to be quite short (see also Table~\ref{tab:2}). The  time dependent dynamics of  QPs is modeled using the Rothwarf-Taylor differential equation for quasiparticles\cite{2014Wang}
\begin{equation}
 \dot{x}_{qp} = -r x_{qp}^{2} -s \,  x_{qp} + g(t).
\end{equation}
Here $s$ is the  trapping rate, $r$ is the recombination rate,  and $g(t)$
is the generation rate from both a constant background and  from the time dependent QP generation pulse. The dynamics of $\delta x_{qp}$ from the QP generation pulse reduces to a single exponential decay if trapping dominates
\begin{equation}
\delta x_{qp} (t) = \delta x^0_{qp} e^{-st}
\label{eq:4}
\end{equation}
where $\delta x^0_{qp}$ is the initially injected QP density immediately after the generation pulse. 
Using eq. (\ref{eq:4}), the trapping rate $s$ at the different applied $\bf B$ is extracted and listed in Table~\ref{tab:2}. We notice both a dramatic increase in $s$ and a large decrease in the  background energy lifetime $T_1$ when the magnetic field $\bf B$ is increased above 23 $\mu$T. This value for $\bf B$ is similar to the estimated  critical field for vortex expulsion $B_m \sim \Phi_0 / w^2 \simeq 21\ \mu$T \cite{2004Stan}, where $\Phi_0 = h/2e$ is the magnetic flux quantum and $w \sim 10 \, \mu m$ is the dimension of the electrodes that form the junction. Therefore, we attribute both the effect on $T_{1}$ and $s$ to the entry of a vortex into one of the arms of the junction electrode. 

\begin{table}[h]
    \caption{Quasiparticle trapping rate $\it{s}$, background energy relaxation time $T_1$, and  charge parity rate $\Gamma_{e \leftrightarrow o}$ at different applied magnetic field $\bf{B}$ for device Q3. At ${\bf B} = 39 \, \mu$T, the quasiparticle recovery time  was too short to be able to accurately  resolve $s$.}
    \label{tab:2}
    \centering
    \begin{tabular}{|c|c|c|c|c|c|c|}
        \hline
        $\bf{B} \, (\mu$T) & 0 & 16 & 23 & 31 & 39 & 47\\\hline
        $T_1 \, (\mu$s) & 82 & 61 & 72 & 5.5 & 13 & 11\\\hline
        {\it s} (ms$^{-1}$) & 6.52 & 4.66 & 4.12 & 138 & & 22.4\\\hline
        $\Gamma_{e \leftrightarrow o}$ (s$^{-1}$) & 144 & 126 & 126 & 123 & 117 & 111\\\hline
    \end{tabular}
\end{table}

Table~\ref{tab:2} also lists the measured charge parity rates $\Gamma_{e \leftrightarrow o}$ at the different magnetic fields. First, we note the measured $\Gamma_{e \leftrightarrow o}$ for these measurements is approximately 4.5 times larger  than $\Gamma_{e \leftrightarrow o}$ measured in Table \ref{tab:1}; this increase of $\Gamma_{e \leftrightarrow o}$ is attributed to the removal of an inner magnetic shielding can between the two experiments, resulting in potentially less shielding of higher temperature blackbody radiation. Second, while the applied magnetic field is able to effectively trap generated excess quasiparticles, only a small decrease in $\Gamma_{e \leftrightarrow o}$ is observed at the larger $\bf{B}$ fields. Therefore, a majority of the observed parity switching events is attributed to production of QPs across the junction arising from absorption of pair-breaking photons. In this model, unwanted pair-breaking radiation couples to the qubit via an antenna mode of the device which is then converted into quasiparticles at the junction \cite{2021Rafferty,2024Liu}.  \par

\begin{figure}[h]
    \centering
    \includegraphics[width=\linewidth]{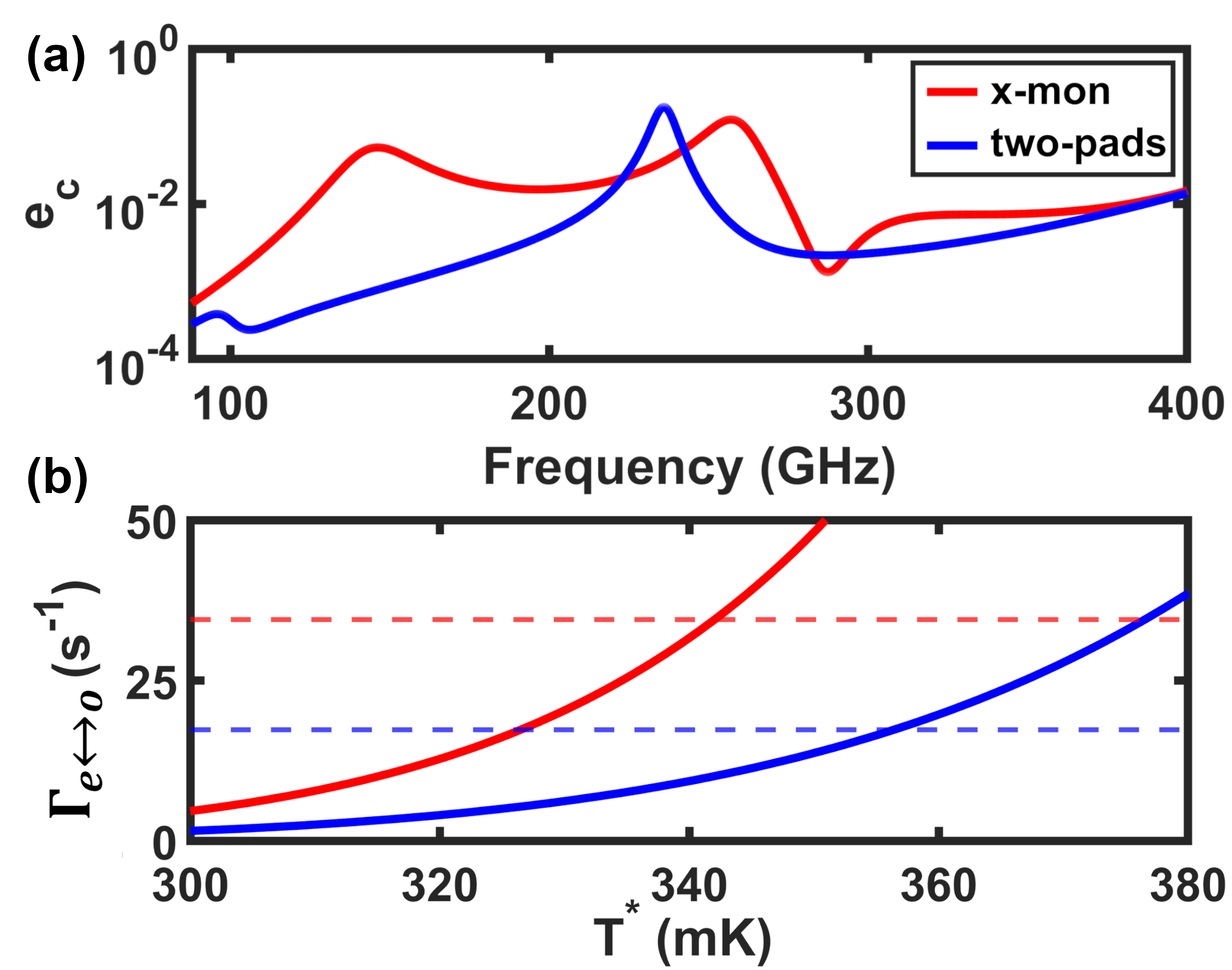}
    \caption{(a) Simulated coupling efficiency $e_c$ and (b) charge parity rate $\Gamma_{e \leftrightarrow o}$ arising from pair-breaking photon absorption for the x-mon (red) and two-pads (blue) designs. The experimentally measured $\Gamma_{e \leftrightarrow o}$ is indicated by dashed lines.}
    \label{fig:4}
\end{figure}

To quantitatively model this rectification effect we follow Rafferty \textit{et al.} \cite{2021Rafferty} and calculate the power efficiency that pair-breaking radiation couples and is then transferred across the junction. First, the radiation impedance $Z_{rad}(f)$ of the shunting capacitors for the x-mon and two-pads geometry was simulated using High Frequency Software Simulator (HFSS). The admittance of the junction was included using a resistively and capacitively shunted junction (RCSJ) model where $Y_j = 1/Z_j=1/R_n + i 2\pi f C_j$, $R_n$ is the normal state resistance, and $C_j$ is the junction self-capacitance. Fig. \ref{fig:4}a. shows the simulated coupling efficiency of the JJ to free space\cite{1965Kurokawa,2021Rafferty}
\begin{equation}
e_c(f) = 1-\left|\frac{Z_{rad}-Z^*_j}{Z_{rad}+Z_j}\right|^2 
\end{equation}
for the x-mon and the two-pads geometries and for pair-breaking frequencies  2$\Delta_{Al,-\mkern-8mu\times\mkern-8mu-} \, / \, h \leq f < 400$ GHz. The x-mon geometry exhibits better impedance matching from free space to the junction between frequencies of 100 GHz to 200 GHz, which results in an overall larger $\Gamma_{e \leftrightarrow o}$ compared to our two-pads geometry.
 
 Finally, to estimate the total photon-assisted generation rate of QPs across the junction, we assume a single-mode background blackbody radiation at an effective temperature $T^*$ as the source. The charge parity rate predicted by this model is
\begin{equation}
\Gamma_{e \leftrightarrow o} = \int_{2\Delta_{Al,-\mkern-8mu\times\mkern-8mu-}} \frac{e_c}{e^{hf/k_B T^*}-1} \, df.
\end{equation}
Fig. \ref{fig:4}b shows the calculated $\Gamma_{e \leftrightarrow o}$ as a function of $T^*$. It is clear that the x-mon geometry has an overall larger integrated power efficiency making it more susceptible to pair breaking radiation. The measured $\Gamma_{e \leftrightarrow o}$ for the x-mon (red) and two-pads (blue) are plotted as dashed lines in Fig. \ref{fig:4}b, which correspond to effective blackbody temperatures of $T^{*} =$ 342 mK and 356 mK, respectively, very similar to the values reported by Liu {\it et al.} \cite{2024Liu}.

\section{Conclusion}
We have measured the single electron charge parity rate $\Gamma_{e \leftrightarrow o}$ in transmons with different shunting capacitors: an x-mon  and a two-pads geometry and consisting of two  different superconducting materials, Al with $\Delta_{Al,\dashv \, \vdash} \sim 170\ \mu$eV and Ta with $\Delta_{Ta,\dashv \, \vdash} \sim 650 \ \mu$eV. Our x-mon qubits had a $\Gamma_{e \leftrightarrow o}$ that was a factor of two larger  than the two-pad geometry. The similarity of $\Gamma_{e \leftrightarrow o}$ for the tantalum and aluminum devices suggests that $\Gamma_{e \leftrightarrow o}$ is  associated with generation  of QPs at the junction. To further test this conjecture, in one of the Al x-mon devices we have also measured $\Gamma_{e\leftrightarrow o}$ after applying a magnetic field to introduce vortices in a junction electrode. The vortices increased the qubit's  relaxation rate $\Gamma_1$ and trapping rate $s$  but had little change on $\Gamma_{e \leftrightarrow o}$. 

To explain the magnitude of $\Gamma_{e\leftrightarrow o}$ in both geometries measured, we have also simulated and modeled the effect of pair-breaking blackbody radiation absorbed by the Josephson junction and resulting in the formation of quasiparticles. The geometry of the shunting capacitors in these simulations is able to explain the factor of two difference in $\Gamma_{e \leftrightarrow o}$. To explain the overall magnitude of $\Gamma_{e \leftrightarrow o}$, the effective temperature of the background blackbody source needs to be  $T^* \sim 340-360$ mK. Two possible sources for this elevated $T^*$ are the lack of light tight shielding around the device and the lack of low-pass filters with a cut-off frequency of $\sim 100$ GHz on the microwave lines. \par

\section*{Acknowledgments}
Qubits Q1-Q4 used in this study were fabricated at MIT Lincoln Laboratory under the program Superconducting Qubits at Lincoln Laboratory (SQUILL). The traveling wave parametric amplifier was also provided by MIT Lincoln Laboratory. We would like to thank Dr. Tamin Tai for manufacturing the magnetic coil used in this experiment and F. C. Wellstood, S. K. Dutta and C. J. Lobb for useful discussions during the course of this work.

{\appendices
\section*{Device Fabrication}
Qubits Q1-Q4 were fabricated by the SQUILL Foundry at MIT Lincoln Laboratory. For this process, 200 nm of Al was deposited on a high resistivity Si substrate. The Al layer was subsequently etched to create all of the large features including the transmon's shunting capacitor except the Josephson junction. The Al/AlOx/Al junctions were subsequently added using a Dolan bridge double-angle process. The thickness of the two junction electrodes were 30 nm and 160 nm. \par 
For Q5, 200 nm of tantalum was sputtered by Cryostar on a sapphire substrate. The shunting capacitor, resonators and other large features of the device were  defined by photolithography and then the unwanted Ta was wet etched using Ta etchant. After cleaning the photoresist off the device, the Al/AlOx/Al Josephson junctions were added via electron-beam lithography and using a Manhattan deposition process with a nominal thickness of 35 nm and 70 nm respectively.  

\section*{Measurement Setup}
\begin{figure}[h!]
    \centering
    \includegraphics[width=\linewidth]{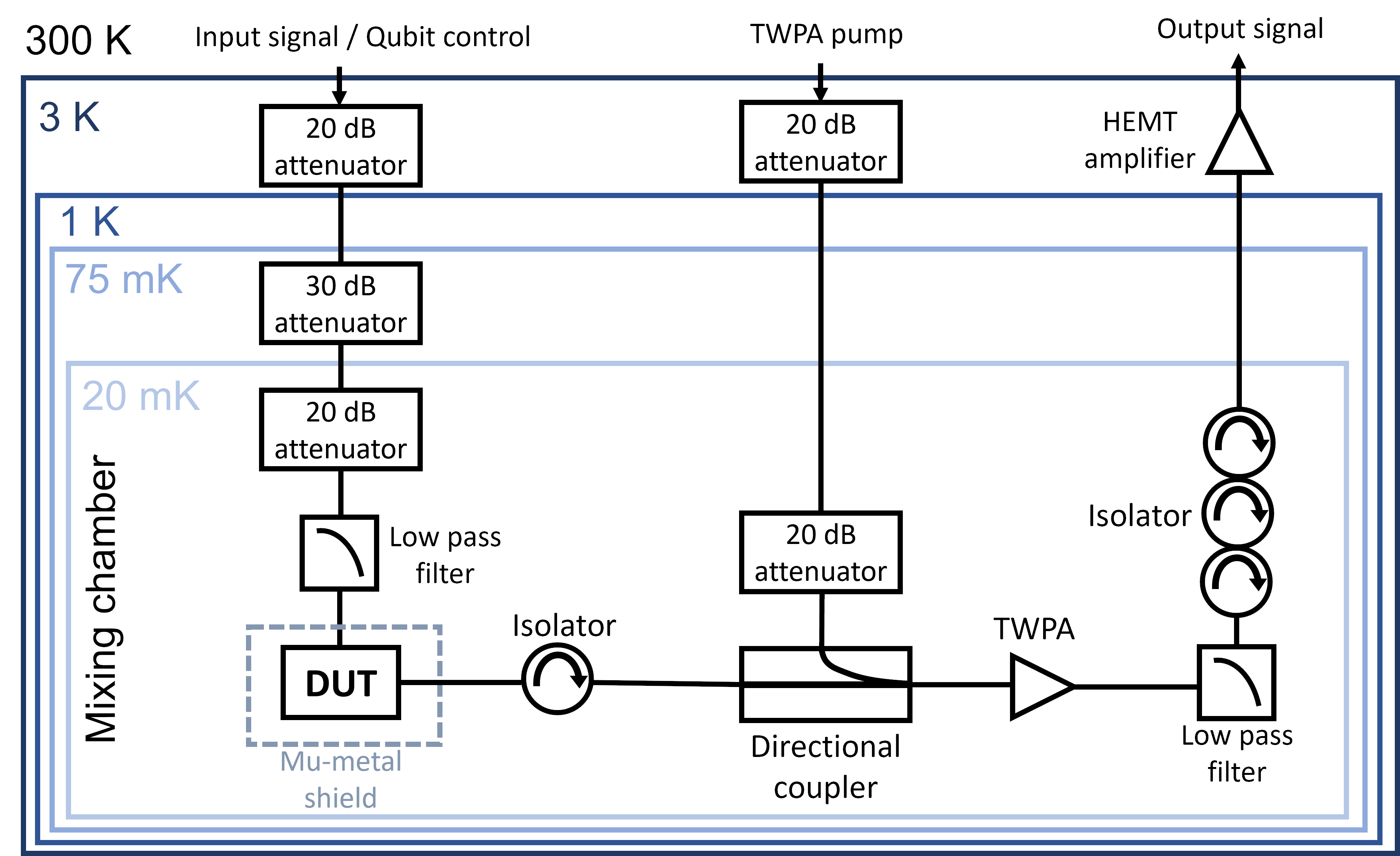}
    \caption{Dilution refrigerator and wiring setup of the microwave measurement. The microwave input and qubit control lines have nominally the same configuration.}
    \label{fig:5}
\end{figure}
The device under test (DUT) consists of a common coplanar waveguide (CPW) transmission line, resonators and qubits. The readout of qubits was dispersively measured at low readout powers by probing the resonator response through the common transmission line. The DUT was mounted to the mixing chamber stage of a cryogen-free Leiden CF-450 dilution refrigerator and cooled to a base temperature of 20 mK. Fig. \ref{fig:5} shows the dilution refrigerator and the microwave measurement setup. For measurements corresponding to Table~\ref{tab:1}, two open ended cylinders of Amumetal 4K surrounded the qubit device. The interior of the inner Amumetal 4K cylinder, which was removed for measurements in Table~\ref{tab:2}, was coated with soot and GE varnish to absorb IR.  Radiation cans made of Cu, Au plated Cu, or Al were bolted to every stage, including the mixing chamber, of the refrigerator. 

For readout and control of the qubit, multiple microwave input lines were attenuated throughout various temperature stages with a total of 70 dB attenuation before it reached the DUT. After the device, a traveling wave parametric amplifier (TWPA)  provided initial signal amplification followed by a high-electron-mobility transistor (HEMT) amplifier at the 3K stage and a low-noise amplifier at room temperature (not shown in figure). We note that IR filters  were not used for these measurements. \par
For experiments that involve applying magnetic field and generating vortices on the device, the innermost Amumetal 4K magnetic shielding cylinder was removed and a home built solenoid coil, with an axial centered on the qubit device, was installed. For each different magnetic field measurement, the mixing chamber was first warmed up to 3K to ensure penetration of the field through the Al package surrounding the qubit device. We then applied an oscillatory field with decreasing magnitude to clear any hysteretic effects. The device was then cooled back down to 20 mK with a newly applied field.
}

\bibliographystyle{IEEEtran}
\bibliography{ref}

\vfill

\end{document}